\documentclass[a4paper,11pt]{article}
\usepackage{pos}

\usepackage{longtable} 
\usepackage{multirow}
\renewcommand{\arraystretch}{1.2} 

\newcommand{\cen}[1]{\multicolumn{1}{c}{#1}}
\newcommand{\ra}[1]{\renewcommand{\arraystretch}{#1}}
\newcommand{\rb}[1]{\renewcommand{\tabcolsep}{#1}}

\definecolor{light-gray}{gray}{0.90}

\title{Searching for New Physics with $B^{0} \to K^{*0} \mu^+ \mu^-$}
\author*[a]{S. Neshatpour}
\author[b]{T. Hurth}
\author[a,c]{F. Mahmoudi}

\affiliation[a]{Universit\'e de Lyon, Universit\'e Claude Bernard Lyon 1, CNRS/IN2P3, \\
Institut de Physique des 2 Infinis de Lyon, UMR 5822, F-69622, Villeurbanne, France}
\affiliation[b]{PRISMA Cluster of Excellence and  Institute for Physics (THEP),\\ 
Johannes Gutenberg University, D-55099 Mainz, Germany}
\affiliation[c]{Theoretical Physics Department, CERN, CH-1211 Geneva 23, Switzerland}
\emailAdd{siavash.neshatpour@univ-lyon1.fr}
\emailAdd{tobias.hurth@cern.ch}
\emailAdd{nazila@cern.ch}

\abstract{
One of the main indications for New Physics in rare $B$-decays is 
deduced from the tension between experimental and Standard Model predictions of 
the angular analysis of the $B^0 \to K^{*0} \mu^+\mu^-$ decay.
There are however possible non-local hadronic effects which in principle can also 
explain these tensions.
In this work,	 we consider a statistical approach for differentiating the source of the tension
in $B^0 \to K^{*0} \mu^+\mu^-$ observables and we also investigate the prospects of such a comparison with
future  data from the LHCb experiment.
}

\FullConference{%
  40th International Conference on High Energy physics - ICHEP2020\\
  July 28 - August 6, 2020\\
  Prague, Czech Republic (virtual meeting)
}

\begin{document}
\maketitle

\section{Introduction}
The tensions between the experimental measurements and the Standard Model (SM) predictions 
of the angular observables in the $B^{0} \to K^{*0}\mu^+\mu^-$ decay~\cite{Aaij:2013qta}
at more than $3\sigma$ was the first in a 
series of deviations  in $b\to  s\ell^+ \ell^-$ transition.
These so-called ``$B$-anomalies'' have also been measured in lepton flavour violating observables $R_K$ and $R_{K^*}$
(with a significance of more than $2\sigma$) for which very precise theoretical predictions 
are available due to cancellation of hadronic quantities. 
Unlike the deviations in the ratios which cannot be explained by underestimated 
hadronic effects, the tensions in the angular observables of $B^0 \to K^{*0} \mu^+ \mu^-$ may
be described by non-local long-distance effects which in principle can mimic $C_9$ (as well as $C_7$) 
New Physics (NP) effects. 
This is readily visible by considering 
that both short-distance NP effects in $C_9$ (and $C_7$),
and non-factorisable hadronic effects contribute to the vectorial helicity amplitude 
\begin{align}
  H_V(\lambda) &=-i\, N^\prime \Big\{ C_9^{\rm eff} \tilde{V}_{\lambda} - C_{9}'  \tilde{V}_{-\lambda}
      + \frac{m_B^2}{q^2} \Big[\frac{2\,\hat m_b}{m_B} (C_{7}^{\rm eff} \tilde{T}_{\lambda} - C_{7}'  \tilde{T}_{-\lambda})
      - 16 \pi^2 {\cal N}_\lambda \Big] \Big\} \,,
\end{align}
where $\lambda=\pm,0$ denotes the helicity of the $K^*$-meson 
and ${\cal N}_\lambda = \big(\text{LO in QCDf} + h_\lambda (q^2) \big)$ 
corresponds to non-factorisable four-quark and chromomagnetic contributions.
While the leading non-factorisable contributions are calculated within
QCD factorisation (QCDf), the higher powers $h_\lambda$ are not calculable within this framework
and are often ``guesstimated''.

The significance of NP fits involving $B^0 \to K^{*0} \mu^+ \mu^-$ observables 
are hence dependent on the assumptions made for the size of the hadronic contributions~\cite{Hurth:2016fbr}.
There have been theoretical calculations of the power corrections
within the LCSR formalism~\cite{Khodjamirian:2010vf} together with exploring the 
analyticity of the amplitude~\cite{Chrzaszcz:2018yza} with recent results~\cite{Gubernari:2020eft} suggesting
a smaller size compared to the previous calculations~\cite{Khodjamirian:2010vf}. 
The power corrections can alternatively be directly fitted to the data~\cite{Ciuchini:2015qxb,Chobanova:2017ghn}.
For the latter approach, we make a statistical comparison with the NP fit to the same data.
However, to be able to make such a comparison, the two scenarios should be embedded~\cite{Chobanova:2017ghn}.
The most general description of unknown power corrections (up to higher order terms in $q^2$)
which respects the analyticity structure of the amplitude is given by~\cite{Arbey:2018ics} 
\begin{align}\label{eq:hlambda}
  h_\pm(q^2)&= h_\pm^{(0)} + \frac{q^2}{1 \,{\rm GeV}^2}h_\pm^{(1)} + \frac{q^4}{1 \,{\rm GeV}^4}h_\pm^{(2)}\,,\\\nonumber
% \label{eq:hlambda0}
h_0(q^2)&= \sqrt{q^2}\times \left( h_0^{(0)} + \frac{q^2}{1\, {\rm GeV}^2}h_0^{(1)} + \frac{q^4}{1\, {\rm GeV}^4}h_0^{(2)}\right)\,,
\end{align}
where considering the nine parameters $h_{\pm,0}^{(0,1,2)}$ to be complex, 
is altogether described by eighteen free parameters.
With such a description, the scenario with contributions to $C_{9}$ is indeed
embedded in this hadronic description. 

We also suggest another description of the power corrections which offers the embedding of  NP contributions to $C_9$ with
a smaller number of parameters~\cite{Hurth:2020rzx}
\begin{align}\label{eq:DeltaC9lambda}
 h_\lambda (q^2)= -\frac{\tilde{V}_\lambda(q^2)}{16 \pi^2} \frac{q^2}{m_B^2}  \Delta C_9^{\lambda,\rm{PC}}\,,
\end{align}
where for each helicity, $\Delta C_9^{\lambda,\rm{PC}}$ is described by one real free parameter 
(or two parameters, if assumed to be complex).
With this three (six) parameter description, there is a better chance 
of getting a meaningful statistical comparison with the NP fit.
Although this minimal description might not be adequate for describing a general 
behaviour of the hadronic contributions, it can capture the distinct behaviours 
for the three different helicities
(unlike the NP contribution which is the same for all three helicities) 
and hence it can be used as a null test for NP.

\section{NP and hadronic fits to recent data on $B^0 \to K^{*0} \mu^+\mu^-$ observables}
We use the latest LHCb angular analysis of $B^0 \to K^{*0} \mu^+\mu^-$ with 4.7 fb$^{-1}$ of data~\cite{Aaij:2020nrf}
together with its branching ratio in the low $q^2$-bins below the charm resonances
as well as the branching ratio of $B\to K^* \gamma$, resulting
in overall 47 observables  (see Ref.~\cite{Hurth:2020rzx} for more details). 
We have used SuperIso 4.1~\cite{Mahmoudi:2007vz} for calculating the observables
resulting in $\chi^2_{\rm SM} = 85.15$.
We first consider NP fits with contribution to  $C_9$ assuming it to be real or complex 
(see Table~\ref{tab:NP_C9_current}) with no uncertainty from power corrections.
\begin{table}[h!]
% \ra{1.}
% \rb{1.3mm}
\begin{center}
\scalebox{0.80}{
%%%%%%%%%%%%%%%%%%%%%%%%%%%%%%%%%%%%%%%%%%%%%%%%
\begin{tabular}{|l|c|c|c|}
\hline
  \multicolumn{2}{|c}{{ $B^{0} \to K^{*0}\, \bar\mu\mu/\gamma$ observables}} & \multicolumn{2}{c|}{($\chi^2_{\rm SM}=85.1$)  }       \\  
\hline
 & \multicolumn{1}{c|}{best-fit value} & $\chi^2_{\rm min}$ & Pull$_{\rm SM}$\\
 \hline \hline
%%%%%%
$\delta C_9$ (real) & $  -1.11	\pm	0.15 $ & $49.7$ &  $6.0\sigma$ \\  
\hline\hline
$\delta C_9$ (complex)& $(-1.04	\pm	0.17) + i(-1.24	\pm	0.61)  $ &  $47.3$ & $5.8\sigma$ \\                                                         
\hline  
\end{tabular}  
}
\caption{One- and two-operator NP fits for real and complex $\delta C_9$, considering $B^{0}\to K^{*0} \bar\mu\mu/\gamma$ observables for $q^2$ bins $\leqslant 8\text{ GeV}^2$.
\label{tab:NP_C9_current}}
\end{center} 
\end{table}  
In the next step we make two hadronic fits considering the 18 parameter description of Eq.~\ref{eq:hlambda}
and the 6 parameter description of Eq.~\ref{eq:DeltaC9lambda}.
In the 18 parameter fit, although the central values of the fitted parameters are non-zero,
they are compatible with zero at the $1\sigma$ level (see left panel of Table~\ref{tab:Had_fit_current}).
This can be understood by the large number of free parameters which cannot be strongly constrained with the current data.
\begin{table}[h!]
\begin{center}
\scalebox{0.77}{
%%%%%%%%%%%%%%%%%%%%%%%%%%%%%%%%%%%%%%%%%%%%%%%%%%%%%
\ra{0.90}
\rb{1.1mm}
\begin{tabular}{|l||r|r|}
\hline
 \multicolumn{3}{|c|}{$B\to K^*\, \bar\mu\mu/\gamma$ observables \iffalse in the low $q^2$ bins up to 8 GeV$^2$ \fi}           \\  
 \multicolumn{3}{|c|}{ ($\chi^2_{\rm SM}=85.15,\; \chi^2_{\rm min}=25.96;\; {\rm Pull}_{\rm SM}=4.7\sigma$)}   \\ 
\hline
% \rowcolor{white}
& \cen{Real}                         & \multicolumn{1}{c|}{Imaginary}   \\ 
 \hline
%%%%%%
$h_{+}^{(0)}$	& $ (	-2.37	\pm	13.50	)\times 10^{-5}	$ & $ (	7.86	\pm	13.79	)\times 10^{-5}	$ \\
$h_{+}^{(1)}$	& $ (	1.09	\pm	1.81	)\times 10^{-4}	$ & $ (	1.58	\pm	1.69	)\times 10^{-4}	$ \\
$h_{+}^{(2)}$	& $ (	-1.10	\pm	2.66	)\times 10^{-5}	$ & $ (	-2.45	\pm	2.51	)\times 10^{-5}	$ \\
\hline											
$h_{-}^{(0)}$	& $ (	1.43	\pm	12.85	)\times 10^{-5}	$ & $ (	-2.34	\pm	3.09	)\times 10^{-4}	$ \\
$h_{-}^{(1)}$	& $ (	-3.99	\pm	8.11	)\times 10^{-5}	$ & $ (	1.44	\pm	2.82	)\times 10^{-4}	$ \\
$h_{-}^{(2)}$	& $ (	2.04	\pm	1.16	)\times 10^{-5}	$ & $ (	-3.25	\pm	3.98	)\times 10^{-5}	$ \\
\hline											
$h_{0}^{(0)}$	& $ (	2.38	\pm	2.43	)\times 10^{-4}	$ & $ (	5.10	\pm	3.18	)\times 10^{-4}	$ \\
$h_{0}^{(1)}$	& $ (	1.40	\pm	1.98	)\times 10^{-4}	$ & $ (	-1.66	\pm	2.41	)\times 10^{-4}	$ \\
$h_{0}^{(2)}$	& $ (	-1.57	\pm	2.43	)\times 10^{-5}	$ & $ (	3.04	\pm	29.87	)\times 10^{-6}	$ \\
\hline
\end{tabular} 
}
\scalebox{0.78}{
%%%%%%%%%%%%%%%%%%%%%%%%%%%%%%%%%%%%%%%%%%%%%%%%%%%%%
\ra{2.}
\rb{1.1mm}
\begin{tabular}{|c|c|}
\hline
 \multicolumn{2}{|c|}{$B \to K^*\, \bar\mu\mu/\gamma$ observables}           \\[-14pt]  
 \multicolumn{2}{|c|}{ ($\chi^2_{\rm SM}=85.15,\; \chi^2_{\rm min}=39.40;\; {\rm Pull}_{\rm SM}=5.5\sigma$)}   \\[-4pt] 
\hline
 & \multicolumn{1}{c|}{best-fit value} \\
 \hline \hline
%%%%%%
$\Delta C_9^{+,{\rm PC}}$ & $\phantom{-}( 3.39	\pm	6.44) + i(-14.98	\pm	8.40)  $   \\                                                         
\hline 
$\Delta C_9^{-,{\rm PC}}$ & $(-1.02	\pm	0.22) + i(-0.68		\pm	0.79)  $   \\                                                         
\hline 
$\Delta C_9^{0,{\rm PC}}$ & $(-0.83	\pm	0.53) + i(-0.89		\pm	0.69)  $   \\                                                         
\hline
\end{tabular}    
}
\caption{Hadronic power correction fit to $B^{0}\to K^{*0}\, \bar\mu\mu/\gamma$ observables 
for $q^2$ bins $\leqslant 8\text{ GeV}^2$, 
considering the 18 parameter description of Eq.~\ref{eq:hlambda} on the left,
and the 6 parameter description of Eq.~\ref{eq:DeltaC9lambda} on the right. 
\label{tab:Had_fit_current}}
\end{center} 
\end{table}
%%
% \clearpage
%%%%
\begin{figure}[h!]
\centering
\includegraphics[width=0.48\textwidth]{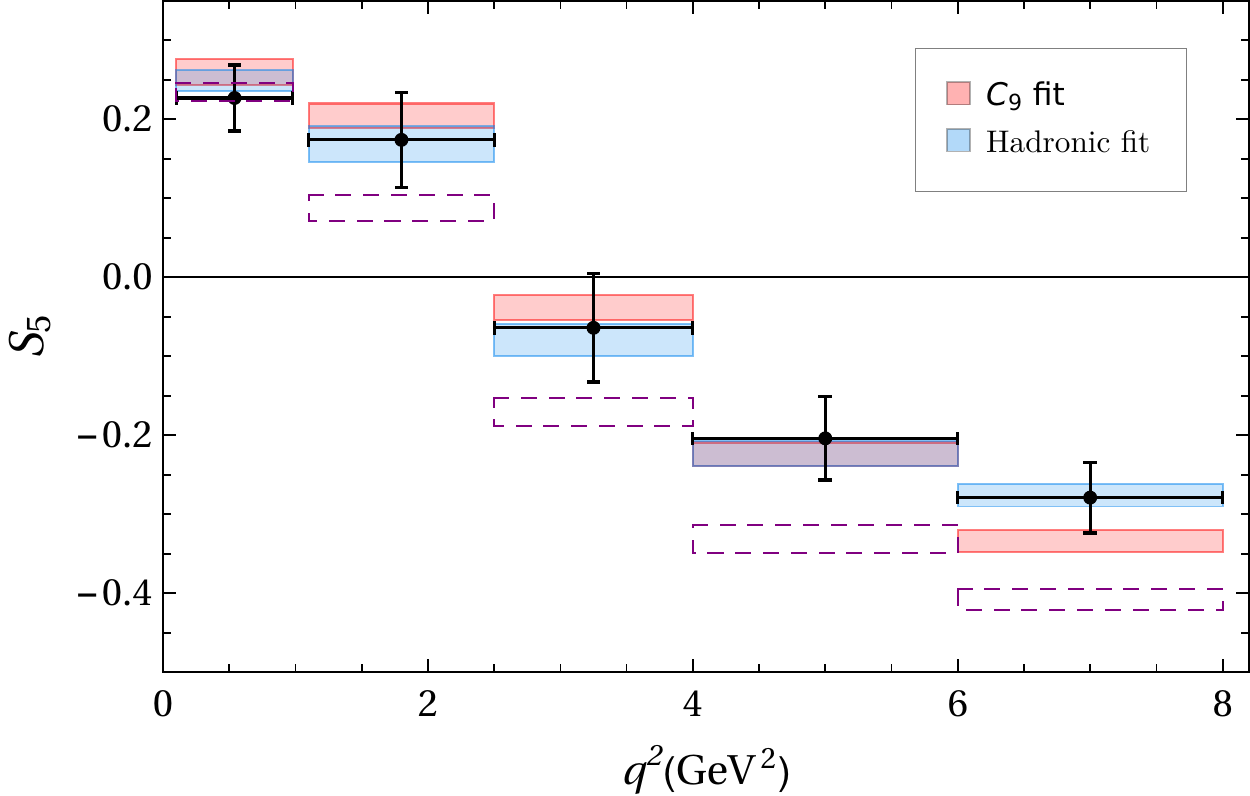}\quad	
\includegraphics[width=0.48\textwidth]{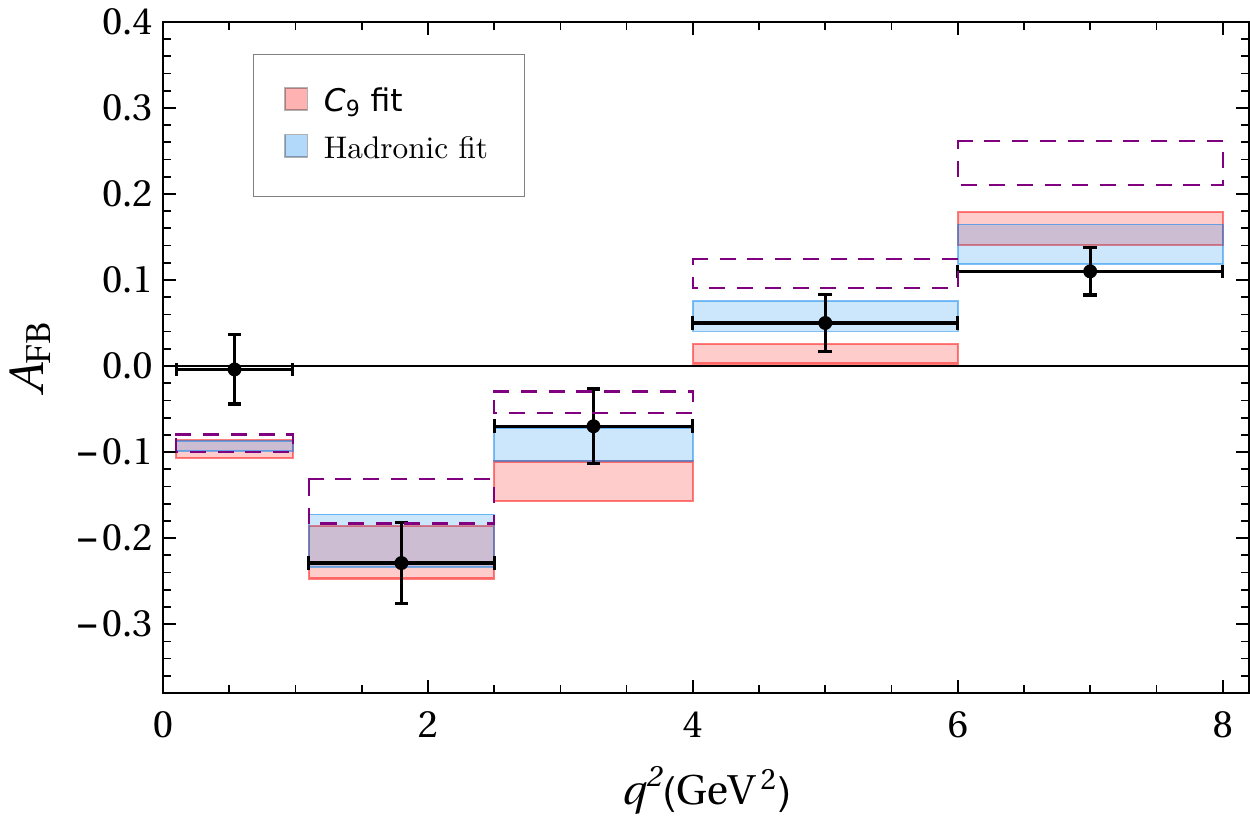}
\caption{NP and hadronic fits at the observable level.
The black crosses show the LHCb measurements~\cite{Aaij:2020nrf} and the the purple dashed boxes correspond to the SM binned predictions.
NP fit to $C_9$ and the hadronic 18 parameter fit are shown with red and blue boxes, respectively.
\label{fig:Lowq2AFBS5}}
\end{figure}
\noindent\hspace{-.5cm}
A potential solution to overcome this is by considering a hadronic description 
with fewer free parameters as given in Eq.~\ref{eq:DeltaC9lambda}. 
Here, the fitted central values for the
three different helicities are not the same (see right panel of Table~\ref{tab:Had_fit_current}), 
hinting that $C_9$ would not be able to offer a similarly good description,
however, the fitted parameters are compatible with each other at $1\sigma$ level and hence
with the current data no conclusive judgment can be made.

The NP and the hadronic fits both give better descriptions of the data compared to the SM,
as can be seen at the observable level in Figure~\ref{fig:Lowq2AFBS5} for $S_5$ and $A_{\rm FB}$. 
Besides the significance of the improvement of each of the fits compared to the 
SM which is at the level of $4.7\sigma$ and more, we can also make statistical comparisons between the NP and the hadronic fits 
as the former is embedded in the latter.
From the second row of Table~\ref{tab:fitComparison}, we can see that by adding 5~(17) more parameters
compared the NP explanation there is only a slight improvement with significances of less than~$2\sigma$. 
This is a strong indication that with the current data, the NP interpretation is a valid option, 
although with the current data, the situation remains inconclusive.

%%%%%Wilks Test - Real & Complex NP%%%%%%%%%%
\begin{table}[h!]
\ra{1.2}
\rb{1.3mm}
\begin{center}
%%%%%%%%%%%%%%%%%%%%%%%%%%%%%%%%%%%%%%%%%%%%%%%%
%%%%%%%%%%%%%%%%%%%%%%%%%%%%%%%%%%%%%%%%%%%%%%%%
\scalebox{0.8}{
\begin{tabular}{|l|c|c|c|c|}
\hline
%%%%%%
\multicolumn{5}{|c|}{$B\to K^*\, \bar\mu\mu/\gamma$ observables;\quad low-$q^2$ bins up to 8 GeV$^2$ }           \\  
\hline
\multicolumn{1}{|c|}{\multirow{2}{*}{nr. of free}} & 1 & 2 & 6 & 18 \\ [-2pt]
\multicolumn{1}{|c|}{parameters} &
$\footnotesize \left(\!\!\begin{array}{c} {\rm Real} \\ \delta C_9 \end{array}\!\!\right)$ &
$\footnotesize \left(\!\!\begin{array}{c} {\rm Comp.} \\ \delta C_9 \end{array}\!\!\right)$ &
$\footnotesize \left(\!\!\begin{array}{c} {\rm Comp.} \\ \Delta C_9^{\lambda,{\rm PC}} \end{array}\!\!\right)$ &
$\footnotesize \left(\!\!\begin{array}{c} {\rm Comp.} \\ h_{+,-,0}^{(0,1,2)} \end{array}\!\!\right)$\\
\hline
0 (plain SM)						&	 $6.0\sigma$	&	 $5.8\sigma$ 	&	 $5.5\sigma$	&	 $4.7\sigma$\\
1 {\small(Real $\delta C_9$)}				&	 $\text{---}$	&	 $1.5\sigma$ 	&	 $1.8\sigma$ 	&	 $1.5\sigma$\\
2 {\small(Comp. $\delta C_9$)}				&	 $\text{---}$	&	 $\text{---}$ 	&	 $1.7\sigma$ 	&	 $1.4\sigma$\\
6 {\small(Comp. $\Delta C_9^{\lambda,{\rm PC}}$)}	&	 $\text{---}$	&	 $\text{---}$ 	&	 $\text{---}$ 	&	 $0.1\sigma$\\
\hline
\end{tabular} 
}
%%%%%%%%%%%%%%%%%%%%%%%%%%%%%%%%%%%%%%%%%%%%%%%%
\caption{Improvements of the NP and hadronic fits compared to the SM and to each other.
\label{tab:fitComparison}
}
\end{center} 
\end{table}  
%%%%%Wilks Test - Real & Complex NP%%%%%%%%%%

\vspace{-1.cm}
\section{Future projections}
We consider future projections of our statistical comparisons for three benchmark cases with an integrated luminosity of 
13.9/fb at the end of Run 2, as well as 50/fb at the end of the first upgrade and finally 300/fb at the end of the second LHC upgrade 
(see Ref.~\cite{Hurth:2020rzx} for further details).
Keeping present central values, the three benchmark cases do not give acceptable fits (with $p$-values$\;\approx 0$).
Instead, we assume two extreme (but equally well-justified) scenarios where we consider the experimental data such that:
\emph{A}) the central value of fit to $C_9$ remains the same  \emph{B}) the central values of the hadronic fit remain the same.

In Table~\ref{tab:Wilks_projections} we give the improvement of the fits compared to the SM and also compare 
the hadronic and NP fits with each other via the Wilks' test. In the left panel, within scenario \emph{A},
by construction, the NP fit has a Pull$_{\rm SM}$ of more than $8\sigma$ significance already with 13.9/fb luminosity.
Since the NP fit is embedded in the hadronic fit we also get very good fits for the latter.
However, there is no improvement compared to the NP description, and looking into the fitted values for 
the 18 parameters,  we can see that the uncertainties of most of them are very large  indicating they are not needed to describe the data. 
On the other hand, in scenario \emph{B} as given in the table on the right, with 13.9/fb luminosity
both the NP and the hadronic fits have large Pull$_{\rm SM}$, 
and while the hadronic fit gives a better description with $4\sigma$ significance, the $p$-value of the NP fit 
is also good and the situation remains inconclusive. It is only with  the 50/fb projection that 
the hadronic description is significantly better than the NP one and also the latter has a very small $p$-value.

%%%%%Wilks Test - Real & Complex NP%%%%%%%%%%
\begin{table}[h!]
\ra{1.2}
\rb{1.3mm}
\begin{center}
%%%%%%%%%%%%%%%%%%%%%%%%%%%%%%%%%%%%%%%%%%%%%%%%
%%%%%%%%%%%%%%%%%%%%%%%%%%%%%%%%%%%%%%%%%%%%%%%%
\scalebox{0.8}{
\begin{tabular}{|l||c|c||c|c||c|c|}
\hline
%%%%%%
\multicolumn{7}{|c|}{Central value of fit to $C_9$ remains the same }           \\  
\hline
luminosity
&\multicolumn{2}{|c||}{13.9 fb$^{-1}$}  
&\multicolumn{2}{|c||}{50 fb$^{-1}$}  
&\multicolumn{2}{|c|}{300 fb$^{-1}$}  \\
% \hline
& $C_9$ & $h_\lambda$ & $C_9$ & $h_\lambda$ & $C_9$ & $h_\lambda$ \\
% \multicolumn{1}{|c||}{\multirow{2}{*}{nr. of free}} & 1 & 18 & 1 & 18 & 1 & 18 \\ [-2pt]
\hline
{\small plain SM}						&	 $8.1\sigma$	&	 $5.1\sigma$ 	&	 $15.1\sigma$	&	 $12.9\sigma$ 	&	 $21.4\sigma$	&	 $19.6\sigma$ 	\\
{\small Real $\delta C_9$}				&	 $\text{---}$	&	 $0.0\sigma$ 	&	 $\text{---}$	&	 $0.0\sigma$ 	&	 $\text{---}$	&	 $0.0\sigma$ 	\\
\hline
\end{tabular} 
}
%%%%%%%%%%%%%%%%%%%%%%%%%%%%%%%%%%%%%%%%%%%%%%%%
%%%%%%%%%%%%%%%%%%%%%%%%%%%%%%%%%%%%%%%%%%%%%%%%
\scalebox{0.8}{
\begin{tabular}{|l||c|c||c|c||c|c|}
\hline
%%%%%%
\multicolumn{7}{|c|}{Central values of the hadronic fit remain the same }           \\  
\hline
luminosity
&\multicolumn{2}{|c||}{13.9 fb$^{-1}$}  
&\multicolumn{2}{|c||}{50 fb$^{-1}$}  
&\multicolumn{2}{|c|}{300 fb$^{-1}$}  \\
% \hline
& $C_9$ & $h_\lambda$ & $C_9$ & $h_\lambda$ & $C_9$ & $h_\lambda$ \\
% \multicolumn{1}{|c||}{\multirow{2}{*}{nr. of free}} & 1 & 18 & 1 & 18 & 1 & 18 \\ [-2pt]
\hline
{\small plain SM}					&	 $7.9\sigma$	&	 $7.9\sigma$ 	&	 $14.6\sigma$	&	 $22.5\sigma$ 	&	 $18.9\sigma$	&	 $41.8\sigma$ 	\\
{\small Real $\delta C_9$}				&	 $\text{---}$	&	 $4.0\sigma$ 	&	 $\text{---}$	&	 $17.5\sigma$ 	&	 $\text{---}$	&	 $37.4\sigma$ 	\\
\hline
\end{tabular} 
}
%%%%%%%%%%%%%%%%%%%%%%%%%%%%%%%%%%%%%%%%%%%%%%%%
\caption{Prospect of improvements of the NP and hadronic fits compared to the SM and to each other.
On the left we have the scenario where current $C_9$ fit to $B\to K^* \mu^+ \mu^-$ remain the same and 
on the right we have considered the scenario where the 18 parameter hadronic fit  remain the same for the experimental projections.
\label{tab:Wilks_projections}
}
\end{center} 
\end{table}  
%%%%%Wilks Test - Real & Complex NP%%%%%%%%%%

\begin{figure}[!th]
\begin{center}
\includegraphics[width=0.32\textwidth]{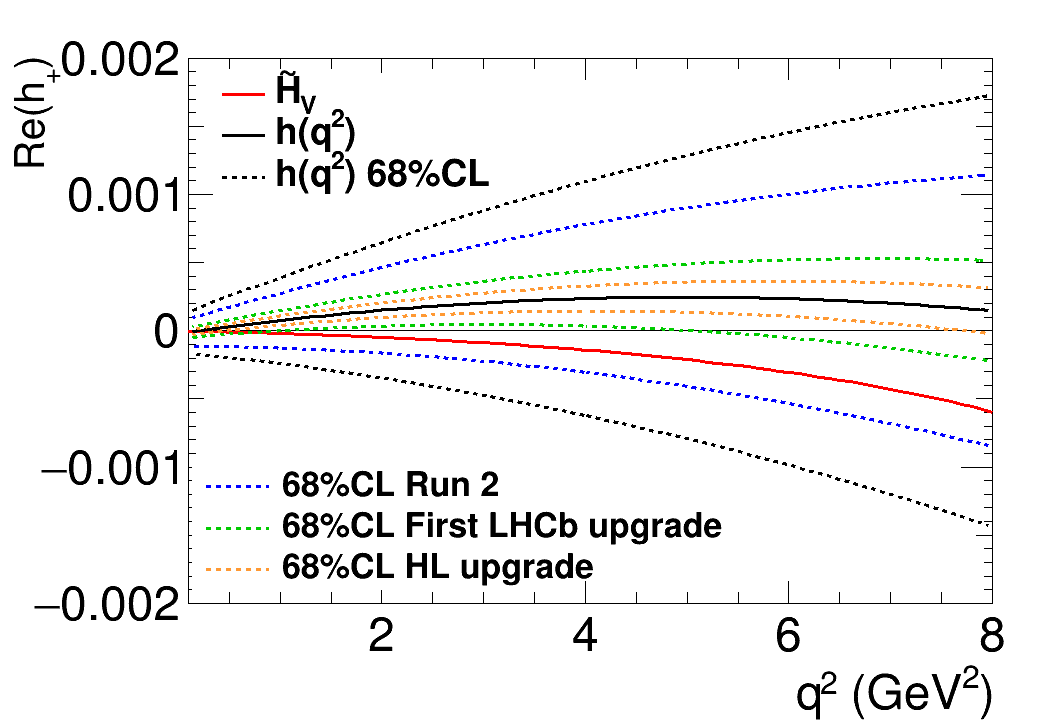}
\includegraphics[width=0.32\textwidth]{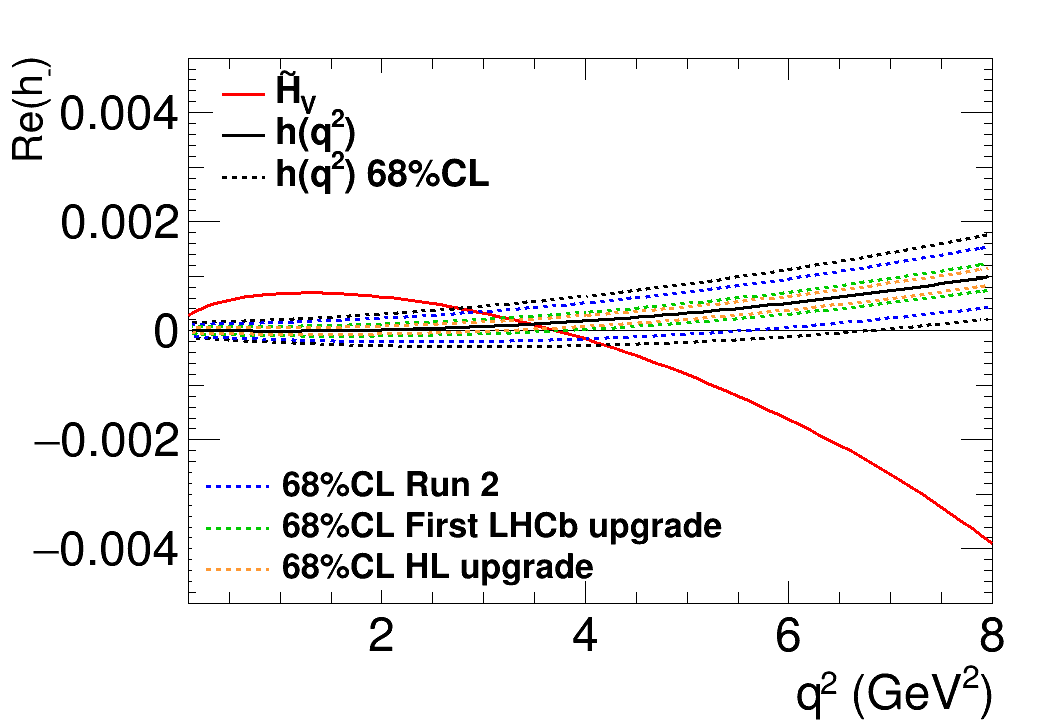}
\includegraphics[width=0.32\textwidth]{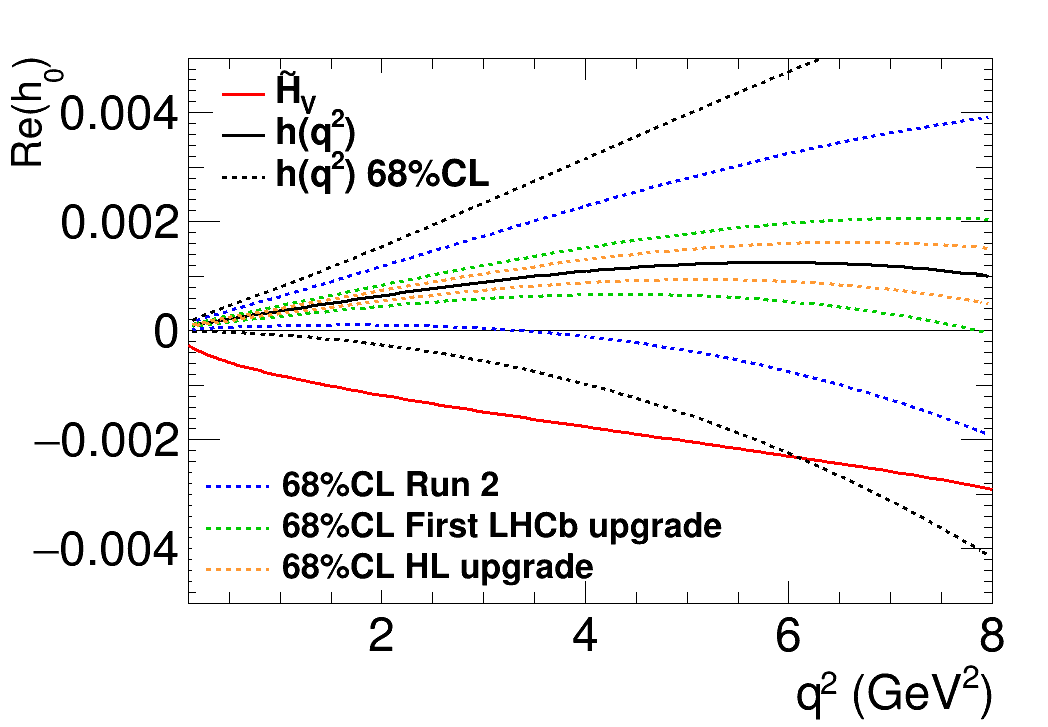}
\caption{Projections for the fitted Re$(h_{\pm,0})$ parameters in scenario \emph{B}.
The solid red lines correspond to the LO QCDf contributions and the 
solid black lines correspond to the central value of the hadronic fit with the current data.
The dashed black, blue, green, and yellow lines correspond to the
68\% C.L. region for current LHCb data, Run 2,  first LHCb upgrade, and second LHCb upgrade, respectively.
\label{fig:hadronic_projections}}
\end{center}
\end{figure}

The central value and the corresponding 68\% confidence level regions of the  hadronic fit projections for the three
benchmark cases of scenario \emph{B} are shown in Figure~\ref{fig:hadronic_projections}. 
It is only after the first LHCb upgrade that the
fitted parameters are no longer consistent with zero.

\section{Conclusions}
A data-driven approach to clarify whether the source of tensions in 
$B^{0} \to K^{*0} \mu^+\mu^-$ is due to underestimated hadronic corrections or genuine New Physics effect is
by considering the statistical comparison between fits of these two explanations to the data.
In this study, we used two statistical tests to make comparisons 
and found out that while with the current data a conclusive judgment is not possible, 
with future data there is a good chance of disentangling the source of the anomalies, 
especially after the first LHCb upgrade.


\begin{thebibliography}{99}

\bibitem{Aaij:2013qta}
LHCb collaboration, R.~Aaij et~al., 
  \emph{{Measurement of
  Form-Factor-Independent Observables in the Decay $B^{0} \to K^{*0} \mu^+
  \mu^-$}}, 
  \href{http://dx.doi.org/10.1103/PhysRevLett.111.191801}{\emph{Phys.
  Rev. Lett.} {\bf 111} (2013) 191801},
  [\href{http://arxiv.org/abs/1308.1707}{{\tt arXiv:1308.1707}}].

\bibitem{Hurth:2016fbr}
T.~Hurth, F.~Mahmoudi and S.~Neshatpour, 
\emph{{On the anomalies in the latest LHCb data}},
  \href{http://dx.doi.org/10.1016/j.nuclphysb.2016.05.022}{\emph{Nucl. Phys.}
  {\bf B909} (2016) 737--777}, [\href{http://arxiv.org/abs/1603.00865}{{\tt arXiv:1603.00865}}].  
  
\bibitem{Khodjamirian:2010vf}
A.~Khodjamirian, T.~Mannel, A.~A. Pivovarov and Y.~M. Wang, 
  \emph{{Charm-loop
  effect in $B \to K^{(*)} \ell^{+} \ell^{-}$ and $B\to K^*\gamma$}},
  \href{http://dx.doi.org/10.1007/JHEP09(2010)089}{\emph{JHEP} {\bf 09} (2010)
  089}, [\href{http://arxiv.org/abs/1006.4945}{{\tt arXiv:1006.4945}}].

\bibitem{Chrzaszcz:2018yza}
M.~Chrzaszcz, A.~Mauri, N.~Serra, R.~Silva~Coutinho and D.~van Dyk,
  \emph{{Prospects for disentangling long- and short-distance effects in the decays $B\to K^* \mu^+\mu^-$}},
  \href{http://dx.doi.org/10.1007/JHEP10(2019)236}{\emph{JHEP} {\bf 10} (2019)
  236}, [\href{http://arxiv.org/abs/1805.06378}{{\tt arXiv:1805.06378}}].
  
\bibitem{Gubernari:2020eft}
N.~Gubernari, D.~van Dyk and J.~Virto,
``Non-local matrix elements in $B_{(s)}\to \{K^{(*)},\phi\}\ell^+\ell^-$,''
[\href{http://arxiv.org/abs/2011.09813}{{\tt arXiv:2011.09813}}].

\bibitem{Ciuchini:2015qxb}
M.~Ciuchini, M.~Fedele, E.~Franco, S.~Mishima, A.~Paul, L.~Silvestrini and M.~Valli,
\emph{{$B\to K^* \ell^+ \ell^-$ decays at large recoil in the Standard Model: a theoretical reappraisal}},
JHEP \textbf{06} (2016), 116
\href{http://doi:10.1007/JHEP06(2016)116 JHEP06(2016)116}{\emph{JHEP} {\bf 06} (2016) 116},
[\href{http://arxiv.org/abs/1512.07157}{{\tt arXiv:1512.07157}}].

\bibitem{Chobanova:2017ghn}
V.~G. Chobanova, T.~Hurth, F.~Mahmoudi, D.~Martinez~Santos and S.~Neshatpour,
  \emph{{Large hadronic power corrections or new physics in the rare decay $B \to K^* \mu^+ \mu^-$?}},
  \href{http://dx.doi.org/10.1007/JHEP07(2017)025}{\emph{JHEP} {\bf 07} (2017)
  025}, [\href{http://arxiv.org/abs/1702.02234}{{\tt arXiv:1702.02234}}].

\bibitem{Arbey:2018ics}
A.~Arbey, T.~Hurth, F.~Mahmoudi and S.~Neshatpour, 
  \emph{{Hadronic and New
  Physics Contributions to $b \to s$ Transitions}},
  \href{http://dx.doi.org/10.1103/PhysRevD.98.095027}{\emph{Phys. Rev.} {\bf
  D98} (2018) 095027}, [\href{http://arxiv.org/abs/1806.02791}{{\tt
  arXiv:1806.02791}}].

\bibitem{Hurth:2020rzx}
T.~Hurth, F.~Mahmoudi and S.~Neshatpour,
``Implications of the new LHCb angular analysis of $B \to K^* \mu^+ \mu^-$ : Hadronic effects or new physics?,''
\href{http://dx.doi.org/10.1103/PhysRevD.102.055001}{\emph{Phys. Rev.} {\bf
D102} (2020) 055001}, [\href{http://arxiv.org/abs/2006.04213}{{\tt
arXiv:2006.04213}}].


\bibitem{Aaij:2020nrf}
LHCb collaboration, R.~Aaij et~al., 
  \emph{{Measurement of
  $C\!P$-averaged observables in the $B^{0}\rightarrow K^{*0}\mu^{+}\mu^{-}$
  decay}},  
  \href{http://dx.doi.org/10.1103/PhysRevLett.125.011802}{\emph{Phys. Rev.
  Lett.} {\bf 125} (2020) 011802}, [\href{http://arxiv.org/abs/2003.04831}{{\tt arXiv:2003.04831}}].

\bibitem{Mahmoudi:2007vz}
F.~Mahmoudi, 
\emph{{SuperIso: A Program for calculating the isospin asymmetry of $B \to K^* \gamma$ in the MSSM}},
  \href{http://dx.doi.org/10.1016/j.cpc.2007.12.006}{\emph{Comput. Phys. Commun.} {\bf 178} (2008) 745--754},
  [\href{http://arxiv.org/abs/0710.2067}{{\tt arXiv:0710.2067}}];
  F.~Mahmoudi,
\emph{{SuperIso v2.3: A Program for calculating flavor physics observables in Supersymmetry}}, 
\href{http://dx.doi.org/doi:10.1016/j.cpc.2009.02.017}{\emph{Comput. Phys. Commun.} \textbf{180} (2009) 1579--1613}, [\href{http://arxiv.org/abs/0808.3144}{{\tt arXiv:0808.3144}}];
F.~Mahmoudi,
\emph{{SuperIso v3.0, flavor physics observables calculations: Extension to NMSSM}}, 
\href{http://dx.doi.org/doi:10.1016/j.cpc.2009.05.001}{\emph{Comput. Phys. Commun.} \textbf{180} (2009) 1718--1719}.
  
  
\end{thebibliography}
\end{document}